# On the variation of solar coronal rotation using SDO/AIA observations

Jaidev Sharma[1]*, Brajesh Kumar [2], Anil K Malik [1] and Hari Om Vats[3]

[1]*Department of Physics, C.C.S. University, Meerut, 250001, India.*

[2]*Udaipur Solar Observatory, Physical Research Laboratory, Dewali, Badi Road, Udaipur 313004 Rajasthan, India.*

[3]*Space Education and Research Foundation, Ahmedabad 380054, India.*

**ABSTRACT**

We report on the variability of rotation periods of solar coronal layers with respect to temperature (or, height). For this purpose, we have used the observations from Atmospheric Imaging Assembly (AIA) telescope on board Solar Dynamics Observatory (SDO) space mission of National Aeronautics and Space Administration (NASA). The images used are at the wavelengths 94 Å, 131 Å, 171 Å, 193 Å, 211 Å, and 335 Å during the period from 2012 to 2018. Analysis of solar full disk images obtained at these wavelengths by AIA is carried out using flux modulation method. Seventeen rectangular strips/bins at equal interval of 10 degrees (extending from $80^0$S to $80^0$N) are selected to extract a time series of extreme ultraviolet (EUV) intensity variations to obtain autocorrelation coefficient. The peak of Gaussian fit to first secondary maxima in the autocorrelogram gives synodic rotation period. Our analysis shows the differential rotation with respect to latitude as well as temperature (or, height). In the present study, we find that the sidereal rotation periods of different coronal layers decrease with increasing temperature (or, height).Average sidereal rotation period at the lowest temperature (~ 600000 Kelvin) corresponding to AIA-171 Å which originates from the upper transition region/quiet corona is 27.03 days. The sidereal rotation period decreases with temperature (or, height) to 25.47 days at the higher temperature (~ 10 million Kelvin) corresponding to the flaring regions of solar corona as seen in AIA-131 Å observations.

**Key words:** Sun: corona, Sun: UV rotation, Sun: rotation

## 1 INTRODUCTION

The rotational profile of the solar corona has created a great interest for the scientific community due to its latitudinal as well as altitudinal variations and thereby illustrating a migration from rigid to differential nature. This phenomenon is not well understood because of less prominent features in the spatial and temporal extent of the corona as compared to the solar photosphere. During the last decade, results show that coronal rotation period varies as low to high values from equator towards the poles. Tracer tracking method is one of the oldest method in which features on solar full disc (SFD) image like sunspots, granules, faculae and coronal bright points (CBPs) etc. are tracked to obtain solar rotation period (Newton and Nunn 1951; Howard et al.1984; Balthasar et al.1986). The various reports, viz., Howard (1991, 1996), Shivraman et al. (1993) showed that all the features such as sunspots (SSNs), plages, filaments, faculae, coronal bright points (CBPs), supergranules, coronal holes, giant cells, etc. on the solar full disc (SFD) images also rotate as that of the Sun. Its rotational profile with respect to latitude as well as altitude/height can be obtained by tracing the passage of aforementioned features across the SFD.

Various studies on the latitude dependent rotational profiles of solar corona by using soft X-Ray (SXR) data (Timothy et al. 1975; Kozuka et al. 1994) have reported the rigid and differential nature; however, the picture is not so clear until now. For example, study of Weberand Sturrock (2002) using the data obtained from the Yohkoh/SXT shows that coronal rotation has more rigid profile in comparison to photosphere and chromosphere. Kariyappa (2008) studied coronal rotation rate employing SFD images from the soft X-ray telescope (SXT) onboard the Yohkoh and the Hinode solar space missions. They showed that the coronal rotation has differential nature during the solar magnetic cycle as that of its neighbouring lower atmospheric layers; however it is almost independent of the phases of the solar magnetic cycle. Chandra and Vats (2009) performed time series analysis on the latitude blocks of SFD images taken by Nobeyama Radioheliograph (NoRH) telescope at 17 GHz during the period 1990-2001. They obtained differential rotation period with respect to latitude extending from $60^0$ S to $60^0$ N, which are correlated in the phase with respect to the annual sunspot numbers and showed that the differential gradient is in antiphase with the annual sunspot numbers. Their reports show that equatorial rotation rates obtained from aforementioned analysis are in good agreement with rotation of photospheric sunspot regions estimated by Balthasar *et al.*(1986), Howard *et al.* (1984), Pulkkinen and Tuominen (1998), with chromospheric level reported by Brajsa *et al.*(2004) and Karachik *et al.*(2006). Chandra and Vats (2010) further reported that the equatorial rotation period follows a systematic trend as that of sunspot numbers and rotation period depends on the phases of solar activity cycle in case of SFD images between $80^0$S to $80^0$N, obtained from the soft X-ray telescope (SXT) on board the Yohkoh space mission. There are two major extensive works (Altrock et al. 2003, Vats et al.2001) for the estimation of coronal rotation and its variations, one is the optical observations at Fe X and Fe XIV lines almost over three decades and other is the estimation of radio observations at 2.8 GHz as well as other radio frequencies. Altrock et al. (2003) used the observations of Fe X (averaged over 18 years) and Fe XIV (averaged over 26 years) at a radius of 1.15 $R_\odot$ and reported that equatorial coronal rotation period is in good agreement with the rotation of photosphere but at high latitudes, these rotation data differ.

Vats *et al*. (2001) suggested that the localized emission originates from a subpart of the solar disc. There is a possibility that feeble type III burst-like activities occur in the same region of the solar atmosphere. They used model of Aschwanden and Benz (1995) to find the height of radio emission in the solar corona and reported that radio emissions at 11 radio frequencies from 275 to 2800 MHz were originated from the mean height range of $6\times10^4$ km to $15\times10^4$ km above the photosphere. Their results suggest that the sidereal rotation period at 2800 MHz, emitted from lower corona at about $6\times10^4$ km above the photosphere, is 24.1 days. They also reported that at the lower frequencies, sidereal rotation period decreases with height. Recently, space-based images at higher resolution taken by EUV telescopes comprise of well-defined features on corona and hence these could be very useful in precise measurement of solar rotation period. In this paper, we report on the investigation of latitudinal as well as temperature (or, height) dependent differential rotation profile of different solar coronal layers during the period from 2012 to 2018 using Flux Modulation Method on solar full disk images (SFD) from Atmospheric Imaging Assembly (AIA) telescope on board Solar Dynamics Observatory (SDO) space mission at 94 Å, 131 Å, 171 Å, 193 Å, 211 Å and 335 Å. The organization of the manuscript is as follows. In section II, we discuss the details of the data utilized for this investigation as well as the technique used to extract the EUV flux at different latitudes at equal interval of 10 deg from the solar full disc (SFD) images. In section III, we discuss methods used to calculate the rotation period at different latitudes. Section IV contains the result and discussion about the latitudinal as well as altitudinal variation of rotation periods and trend in rotation periods at different confidence levels. Conclusions are given in the last section.

## 2 OBSERVATIONS AND METHODOLOGY

The Solar Dynamics Observatory (SDO), a National Aeronautics and Space Administration (NASA) solar space mission, has three main instruments namely Atmospheric Imaging Assembly (AIA), The Helioseismic and Magnetic Imager (HMI) and the Extreme Ultraviolet Variability Experiment (EVE) that are designed to obtain images of Sun's atmosphere (photosphere, chromosphere, transition region and corona) aimed to study the dynamical activities taking place in the solar environment. The AIA instrument contains filters with 10 different wavelength bands to identify key aspects of solar activities. The spatial resolution of the instrument is ~ 0.6" per pixel that is approximately twice as much as Hinode/XRT resolution (i.e. ~ 1.032" per pixel) and four times of SOHO/EIT resolution (i.e. ~2.629" per pixel) (Lemen et al. 2012). In our study, we used preliminary observations from AIA instrument at aforementioned wavelengths. The available SFD images in the digital format having size 512×512 is utilized from the web-accessible database at a cadence of one image per day (at almost fixed time), from January 2012 to December 2018, with very less data gaps. However, this negligible gap of flux/intensity is filled by interpolation process.

Here, in this analysis, on SFD images total 17 latitudinal rectangular strips/bins/blocks for $\pm80^0$ on both hemispheres at an interval of $10^0$ latitude are chosen. Each rectangular block contains the width only two pixels and length covers total pixels on SFD at each latitude. All the data sets are categorized into one-year intervals from January 2012 to December 2018. A time series of daily values of EUV flux/intensity variations of multiple layers is generated from mean of available pixels values in particular bin. We obtain autocorrelation coefficients using standard subroutines of Interactive Data Analysis (IDL) software to find out rotation period. Most of the time series of EUV flux at different latitudes contains the information of coronal features that dominate in the investigation of rotation period at that latitude. The coronal structure indicates that large-scale emitters can remain on SFD for several rotation periods (Fisher et al. 1984). Due to long lived features/emitters on SFD, it is reasonable to utilize the autocorrelation analysis to obtain the precious rotation rates. (Hansen et al. 1969; Sime et al. 1989; Vats et al. 1998 and Weber et al. 1999) have demonstrated that autocorrelation analysis is a very precious statistical process to find any prominent emitters present in the observations and thereby determine the coronal rotation. During our analysis, we observed that at some latitude bins, the peaks of autocorrelogram are not too smooth to give rotation period that may be because short-lived features behave like a noise. This could also be due to less periodicity in features, interference of noise due to crosstalk and also due to poor availability of features. In order to obtain high accuracy in synodic rotation period, Gaussian fitted peak value of secondary maxima (first peak of autocorrelogram) is used.

The Gaussian function is defined as

$$y = y_0 + \frac{A}{w\sqrt{\frac{\pi}{2}}} e^{\frac{-2(x-x_0)^2}{w^2}} \qquad (1)$$

where $x_0$ = center of the maxima, $w$ = 2 times the standard deviation of the Gaussian fit, $A$ = area under the curve and $y_0$ = baseline offset.

Recently, parametric and nonparametric methods have been applied to ascertain trends in time series observations. However, scientific community migrated towards the application of nonparametric tests because it is more precious for non-normally distributed and censored data, including missing values. These methods are a little bit influenced by the presence of outliers in the data. The MK test (Mann, 1945; Kendall, 1975) is one of the popular methods to identify the trend in time series data. The MK trend test was first carried out by computing an S statistic as follows:

Suppose $x_1, x_2, \ldots, x_n$ represents n observations where $x_j$ represents the data points at time j, then the Mann-Kendall statistics (S) is given by

$$S = \sum_{k=1}^{n-1} \sum_{j=k+1}^{n} sgn(x_j - x_k) \qquad (2)$$

$$sgn(\theta) = \begin{cases} 1 & if\ \theta > 0 \\ 0 & if\ \theta = 0 \\ 1 & if\ \theta < 0 \end{cases} \qquad (3)$$

Above conditions are utilized by assuming that the sample data are independently and identically distributed. Here, S statistic defined by Eq. (2) has mean and variance as follows (Kendall, 1975).

$$E(S) = 0 \qquad (4)$$

$$Var(S) = \frac{1}{18}[n(n-1)(2n+5) - \sum_{p=1}^{g} t_p(t_p - 1)(2 t_p + 5)] \qquad (5)$$

Where n is sample size, g is the number of tied groups (group having the same value) and $t_p$ is the number of data points in the $p^{th}$ group.

The original MK test can be computed as

$$Z = \begin{cases} \frac{(S-1)}{\sqrt{Var(S)}} & S > 0 \\ 0 & S = 0 \\ \frac{(S+1)}{\sqrt{Var(S)}} & S < 0 \end{cases} \qquad (6)$$

If $-Z_{1-\alpha/2} \leq Z \leq Z_{1-\alpha/2}$ is satisfied then at significance level of $\alpha$, the null hypothesis is accepted with no trend. Failing which, the null hypothesis is rejected while alternative hypothesis is accepted at the same significance level. Now the probability density function f(z) has to be defined to ascertain the significant trend (increasing or decreasing). If a normal distribution has mean 0 and standard deviation 1 than the probability density function is defined as follows:

$$f(z) = \frac{1}{\sqrt{2\pi}} e^{\frac{z^2}{2}} \qquad (7)$$

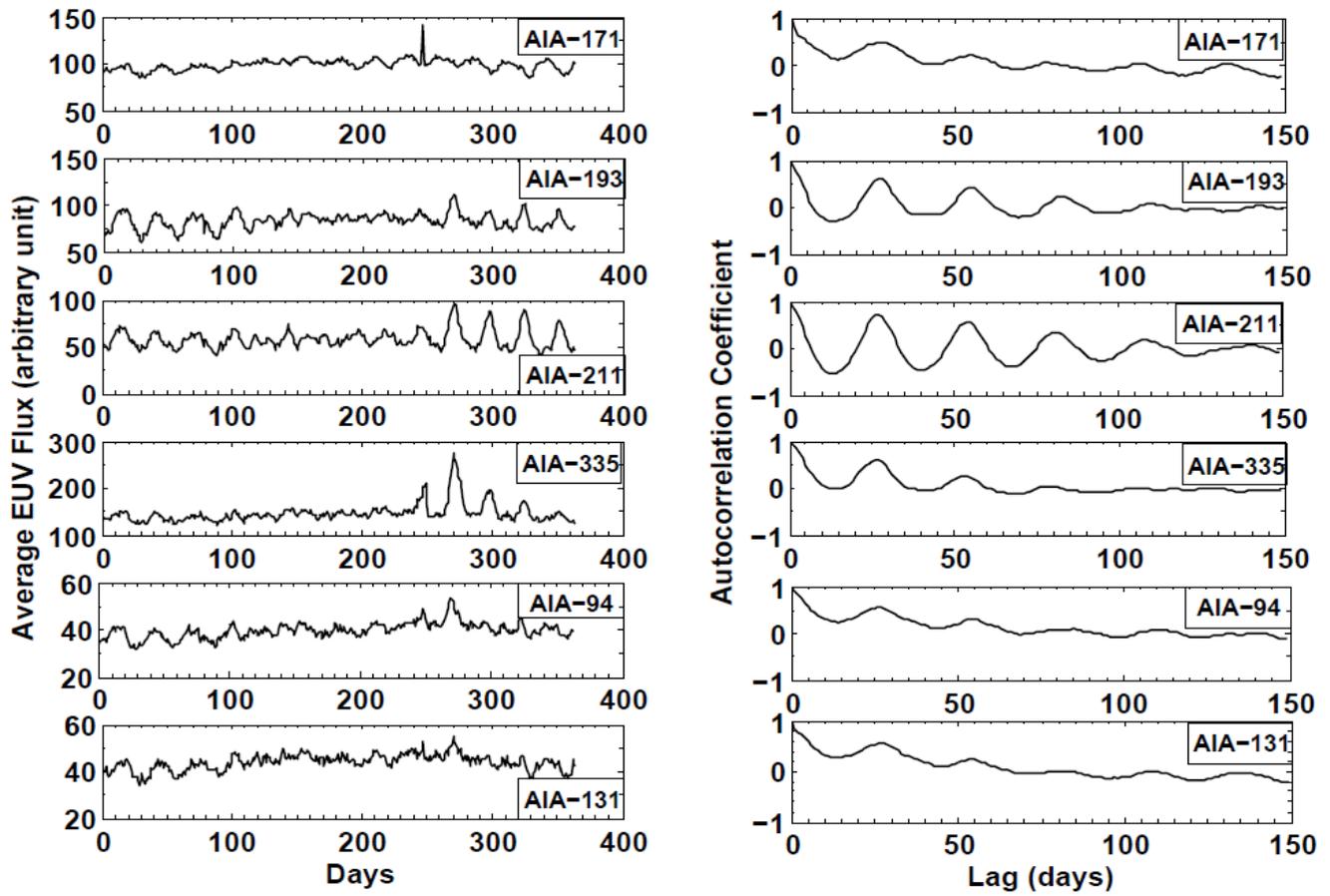

**Figure 1:** Above panels are typical examples of the time series of average EUV flux and corresponding autocorrelograms at $30^0$ South at each wavelength for the year 2017.

If Z is negative having the condition that computed probability has larger value in comparison to the level of significance α then the trend is said to be *decreasing*. On the other hand the trend is said to be *increasing* if Z is positive under the condition that computed probability has larger value as compared to the level of significance α. The condition of *no trend* at significant level α is satisfied if α has larger value than computed probability.

**3 DATA ANALYSIS**

Autocorrelation upto a lag of 150 days is calculated and plotted in the form of autocorrelogram for the time-series of each latitude bins. Figure **1** shows average EUV flux and typical autocorrelograms for a data set of the year 2017 at $30^0$ South at each wavelength. Autocorrelation analysis has been performed by using above mentioned time series flux. Almost all the autocorrelograms obtained by time series flux have smooth nature that causes a fair amount of accuracy in the estimation of rotation period. The autocorrelograms in Figure**1** show several smooth periodic peaks with consecutive decreasing amplitudes. The amplitudes of oscillatory features decrease with increase in lag perhaps due to change in temporal solar rotation period. Here in each case, the first peak of autocorrelogram is fitted by the Gaussian function to estimate the synodic rotation period (location of peak at horizontal axis in days) because it seems to be more clear, smoother and higher than other peaks.

The error in determining synodic rotation period for farthest position of the peak is much greater than for the closest peak (Chandra et.al.**2010**). Thus, in order to obtain the rotational nature at highest accuracy, it appears essential to select the first peak. However, this possible accuracy will vary with latitude bins. Hence, we take the horizontal value (lag in days) of secondary maxima corresponding to first peak of each autocorrelogram to estimate the synodic rotation period. Further to reduce the standard errors and to enhance accuracy, Gaussian fitting is carried out. Standard errors in the fitting of Gaussian function are significantly low which results in high accuracy in the measurements. We have not found sufficient latitudinal coverage of rotation periods in some years (AIA-171 in 2015 and AIA-335 in 2014 & 2018) owing to the less periodicity in data due to uniformity or randomness of EUV emitters or inclusion of noise at corresponding latitude. The data of AIA-335 in year 2016 is not updated at the webpage of SDO so this year is also excluded from the analysis. For the sake of comparison of all findings, the average EUV flux at higher latitudes (>60⁰), in both the hemispheres, have insufficient possibility of containing the rotational periodicity and so is excluded from the analysis. The conversion of synodic rotation period into sidereal rotation period is given by the following equation.

$$T_{sidereal} = \frac{365.26 * T_{synodic}}{365.26 + T_{synodic}} \quad (8)$$

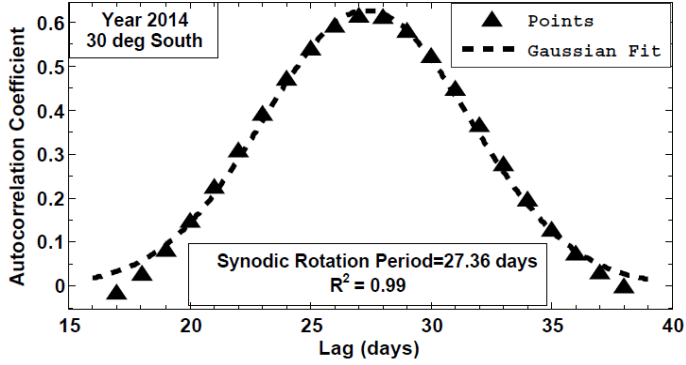

**Figure 2:** The plot shows the Gaussian fit to the 1st secondary maxima of the autocorrelogram of EUV flux at 211 Å, and therefore optimum position of peak needs to be selected.

In previous studies, most of the measurements of solar rotations are traditionally fitted with a polynomial expressing the latitudinal dependence with rotation rate (Howard and Harvey 1970; Schroter et al. 1985) and given as

$$\omega_{rot}(b) = C_0 + C_1 Sin^2 b + C_2 Sin^4 b \quad (9)$$

here $\omega_{rot}(b)$ is solar rotation rate, b is latitude, $C_0$ represents equatorial rotation rate and $C_1$ and $C_2$ represent differential gradients with $C_1$ for lower latitudes and $C_2$ for higher latitudes. In Equation (9), parameters $C_1$ and $C_2$ are hold opposite correlation (termed as cross talk between the coefficients) that causes problems to compare with other results (Snodgrass 1984; Snodgrass and Ulrich 1990). Many techniques to remove this crosstalk have been reported in the literature. For example, Howard et al, 1984; Pulkkinen and Tuominen 1998; Brajsa et al. 2002a; Sudar et al. (2014) put $C_2$=0, Scherrer et al, (1980) put $C_1 = C_2$, and Ulrich et al. (1988) put $C_2 = 1.0216295\ C_1$. However, no justification has been given in favor of these arguments. Vats et al. (2011) reported North–South asymmetry in rotation rate on both the hemispheres. To achieve better accuracy, it is reasonable to fit asymmetric equation to estimate the rotational coefficients. Thus, we fit our data with asymmetric expression given in Equation (10) and observe that our data strongly supports the asymmetric rotational profile.

$$\omega_{rot}(b) = C_0 + C_1 Sinb + C_2 Sin^2 b + C_3 Sin^3 b + C_4 Sin^4 b \quad (10)$$

Here, $C_0$ is equatorial rotation rate, $C_1$ and $C_3$ determine asymmetry on both the hemispheres and $C_2$ and $C_4$ represent the gradients at middle and upper latitudes. The analysis of our data for the period of January 2012 to December 2018 shows that by setting the coefficient $C_4$ =0, fitting process has less root mean square (RMSE) in comparison of other setting ($C_2 = C_4$) for removal of crosstalk. The Sidereal rotation period (in days) corresponding to rotation rate (deg/day) is given as

$$T(b) = \frac{360°}{\omega_{rot}(b)} \quad (11)$$

together with b as latitude.

In Figure 2, autocorrelation coefficient and Gaussian fit of secondary maxima at first peak is plotted for the year 2014 at 30⁰ South. This plot shows that autocorrelation coefficient follows the Gaussian function with most acceptable value of fitting parameter $R^2$ (Pearson's coefficient). Almost similar trend is observed for all latitudes and all the duration considered in this work. The RMSE in fitting process of equation (10) has small values that indicate the fitting is reasonably good with higher accuracy in results.

## 4 RESULTS AND DISCUSSION

The solar rotation profile with respect to height in the solar atmosphere is not clearly understood as yet. Since solar rotation has implications on the solar dynamo processes, it is quite necessary to understand this important phenomenon on the Sun. As of now, it is well known that the solar interior rotates faster as compared to photosphere (Howe 2009). Further; rotation periods of individual latitudes with respect to depth in the solar interior follow a complex pattern. Vats et al. (1999) used radio flux for solar cycles 21 and 22 at 2.8 GHz and determined that the solar corona has higher rotation rate in comparison to lower regions of the Sun. Again Vats *et al.* (2001) used solar radio flux from the Cracow Astronomical Observatory at 275, 405, 670, 810, 925, 1080, 1215, 1350, 1620, 1755 and at 2800 MHz from the Algonquin Radio Observatory in Canada (From duration 1June 1997 to 31 July 1999) and reported that coronal rotation periods have downward trend with respect to increasing height in the coronal layers.

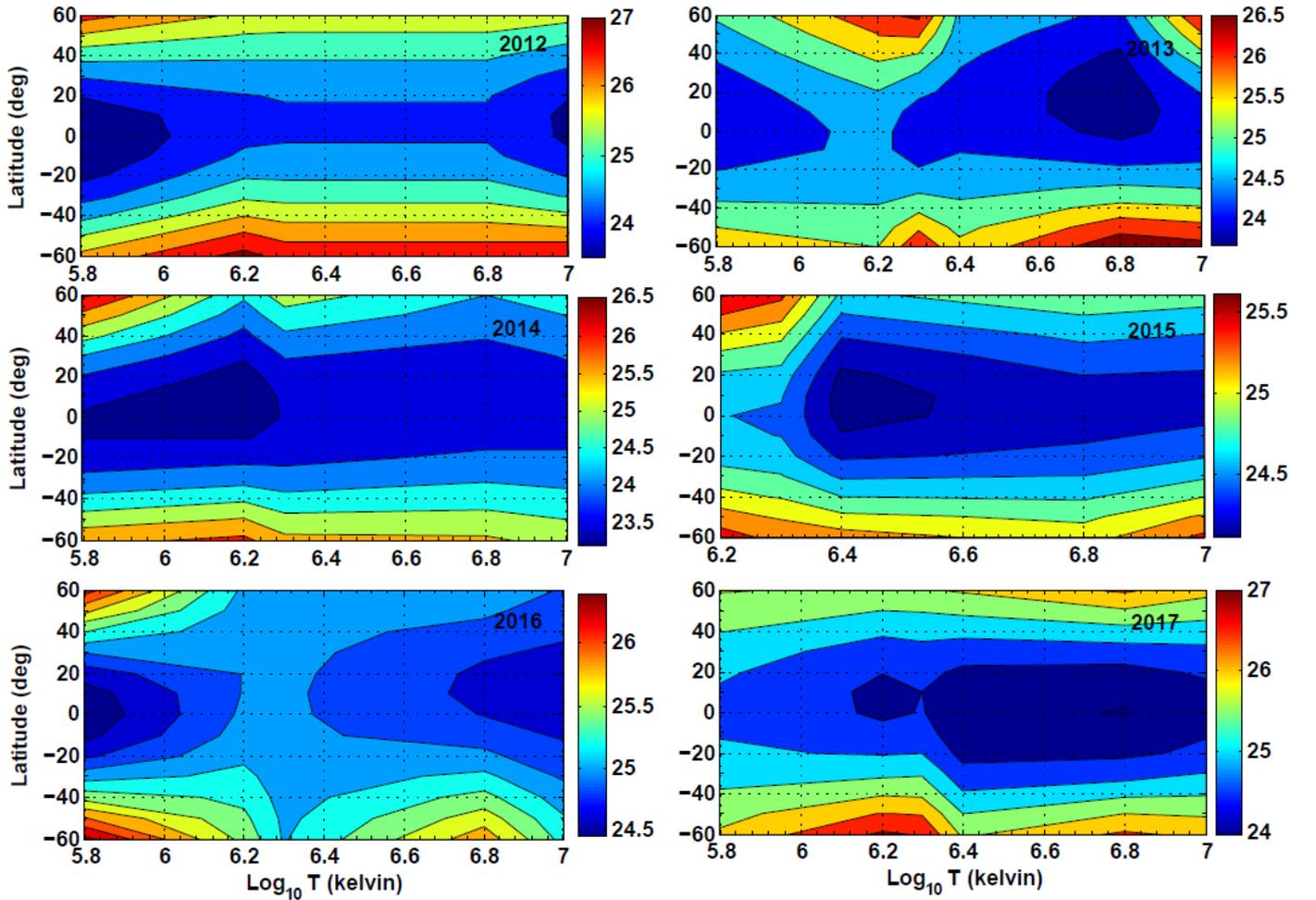

**Figure 3:** Contour plots showing yearly profile of sidereal rotation period in days (shown in color code) from 2012 to 2017 with respect to latitude in degree (vertical direction) as well as temperature in Kelvin in horizontal direction.

## 4.1 LATITUDINAL AS WELL AS ALTITUDINAL PROFILE IN ROTATION

We show in Figure **3** and **4**, the latitude as well as temperature (or, height) dependent profiles of yearly sidereal rotation period for the duration from 2012 to 2018. In these contour plots, vertical direction shows latitudinal rotational profile for each year whereas along the horizontal direction sidereal rotation period with respect to temperature (or, height) is shown.

As expected, these plots show lower sidereal rotation period at equatorial region and these gradually increase towards the poles. Rotation period of individual latitudes (spanning -60 deg to +60 deg) with respect to temperature (or, height) follows a complex pattern as that of solar interior.

## 4.2 TREND IN ROTATION PERIOD

In this analysis, Mann-Kendall test is applied to estimate the significance level in the trend in sidereal rotation period with respect to increasing temperature (or, height). Here, we report yearly trend, overall trend and band wise trend at different confidence levels. Table 1 provides the important parameters of this test like **S**, **Z**, corresponding probability, and trends. These statistical parameters have been discussed in detail in the section 2 of the paper. Present analysis concludes that in most of the years sidereal rotation period decreases with increasing temperature (or, height) at different confidence levels (c.f., Table 2). However, this trend is statistically significant in 2015 and 2017.

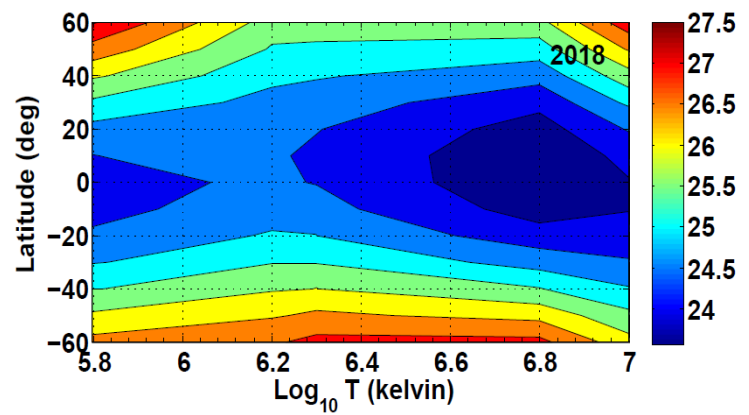

**Figure 4**: Similar as in figure 3 but for the year 2018.

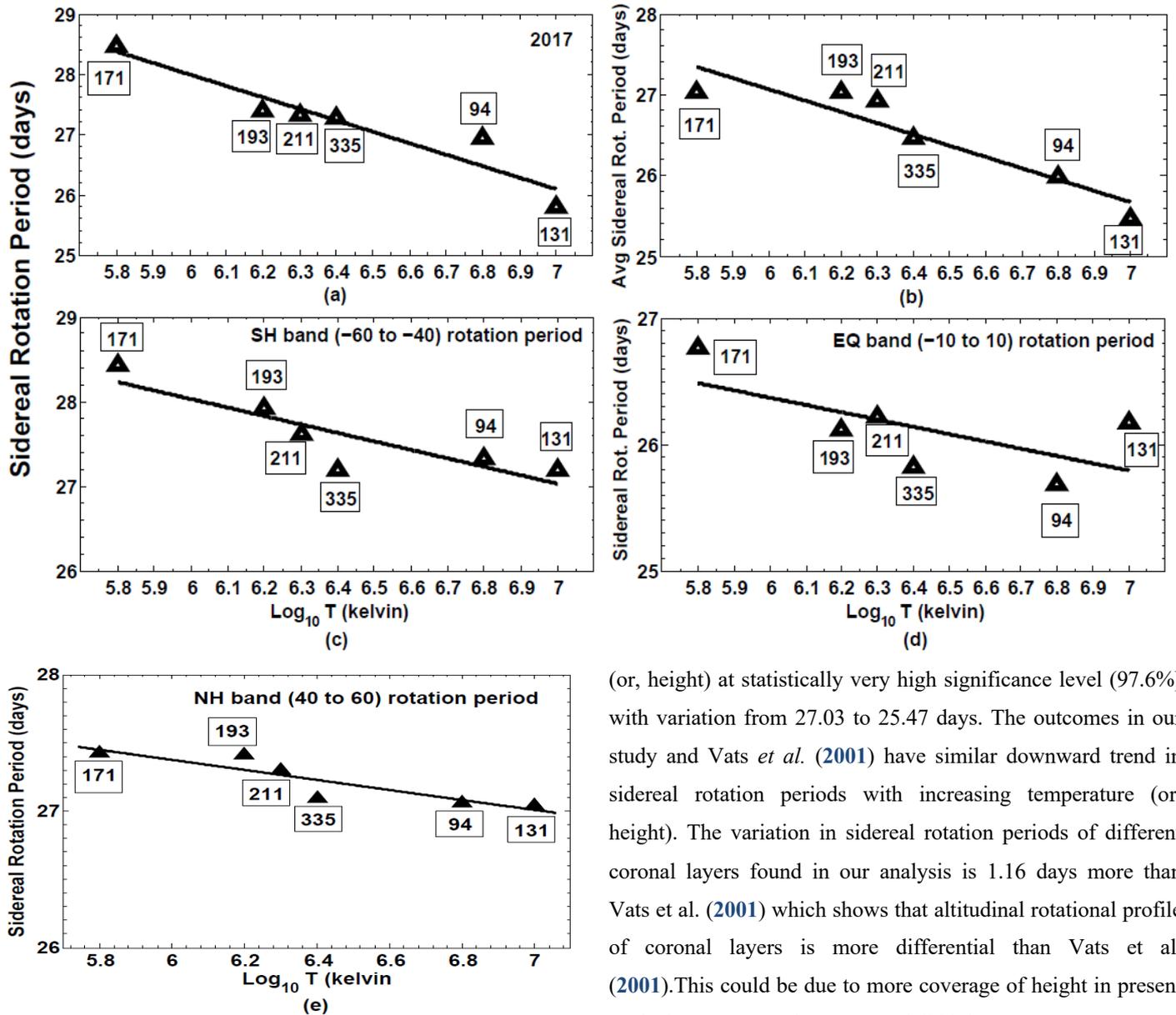

**Figure 5**: Panel (a) shows a typical diagram of downward trend in 2017 whereas panel (b) represents overall trend (yearly averaged) in sidereal rotation period (days) with respect to temperature (kelvin). Panels (c), (d) and (e) represent the trend in the southern hemispheric band (average of -60 to -40 latitude), equatorial band (average of -10 to +10 latitude) and northern hemispheric band (average of 40 to 60 latitude) respectively. In above panels linear regression is carried out to estimate the probable trend. These results show downward trend in sidereal rotation period with increasing temperature (or, height) at different confidence levels (c.f., Table **1**).

The gradients of yearly trends have different values with minimum in 2018 and maximum in 2012 having random systematic variation. Overall trend of sidereal rotation period has interesting downward trend with increasing temperature (or, height) at statistically very high significance level (97.6%) with variation from 27.03 to 25.47 days. The outcomes in our study and Vats *et al.* (**2001**) have similar downward trend in sidereal rotation periods with increasing temperature (or, height). The variation in sidereal rotation periods of different coronal layers found in our analysis is 1.16 days more than Vats et al. (**2001**) which shows that altitudinal rotational profile of coronal layers is more differential than Vats et al. (**2001**).This could be due to more coverage of height in present analysis as compared to Vats et al.(**2001**).

Southern region (-60,-50,-40), equatorial region (-10, 0, 10) and northern region (40, 50, 60) also follow the downward trend in rotation period with increasing temperature (or, height) at considerably good confidence levels (99.7, 86 and 99.2%, respectively) with increasing gradients from South to North directions. Here, variation in rotation periods in this range of temperatures is 4.5%, 2.23% and 1.4%, respectively, that means from Southern region towards Northern region, altitudinal profile of rotation period is consecutively becoming less differential. The reason for this observed phenomenon remains an enigma.

Altogether, our analysis confirms a clear downward trend in rotation period with increasing temperature (or, height) of different coronal layers.

*Table 1: Results of Mann-kendall test*

| Year | S statistic | Z | Probability | Trend |
|---|---|---|---|---|
| 2012 | -5 | -0.751 | 0.548 | Decreasing |
| 2013 | -9 | -1.503 | 0.867 | Decreasing |
| 2014 | -4 | -0.735 | 0.538 | Decreasing |
| **2015** | **-8*** | **-1.715*** | **0.914** | **Decreasing** |
| 2016 | 0 | 0 | 0 | No trend |
| **2017** | **-15*** | **-2.63*** | **0.992** | **Decreasing** |
| 2018 | -15 | -1.225 | 0.780 | Decreasing |
| Avg(2012-18) | -13* | -2.254* | 0.976 | Decreasing |
| **(-60 to -40)** | **-13*** | **-2.939*** | **0.997** | **Decreasing** |
| **(-10 to 10)** | **-7** | **-1.469** | **0.860** | **Decreasing** |
| **(40 to 60)** | **-15*** | **-2.630*** | **0.992** | **Decreasing** |

*\* Statistically significant at 90% confidence level*

*Table 2: Summary of the regions of the solar corona sampled by the various AIA observations at different wavelengths and the corresponding temperatures of those regions (Lemen et al.2012).*

| Channels | Region of solar atmosphere | $Log_{10}(T)$ |
|---|---|---|
| 171Å | quiet corona and upper transition region | 5.8 |
| 193 Å | corona and hot flare plasma | 6.2 |
| 211Å | active-region corona | 6.3 |
| 335Å | active-region corona | 6.4 |
| 94Å | flaring corona | 6.8 |
| 131Å | flaring corona | 7.0 |

## 5 CONCLUSIONS

Our results show that latitude dependent coronal rotation profiles with respect to increasing temperature (or, height) have no systematic variation as observed in the case of solar interior. The reason for this complex pattern is still an enigma. However, average rotation of latitudinal bands (-60,-50,-40), (-10, 0, 10) and (40, 50, 60), follow systematic downward trend in rotation period corresponding to increasing temperature at considerably good confidence levels. The yearly trend in this analysis follows significant downward trend in the years 2015 and 2017, however the remaining years follow this trend at relatively low confidence levels. This could be due to the complexities in the rotational features on coronal layers, noise contents and errors in the tools applied. As far as overall trend in rotation period is concerned, there is a gross downward trend in the solar corona with respect to increasing temperature (or, height). The present data set has six AIA observing channels, so MK test is applied to six data points which is giving reasonably consistent indication of trend. However, in statistical analysis, more number of data points is always recommended for better accuracy. Our findings of rotation rate variation as a function of temperature is reasonably supported by Livingston et al. (1969) who compared rotation in chromosphere and photosphere. They stated that the chromosphere rotates 8% faster than the photosphere. Thus, rotation rate increases (or period decreases) with increasing temperature from photosphere to chromosphere. Similarly, Howe et al. (2009) reported a considerable decreasing trend of average rotation rate from interior of the Sun to outward to the photosphere. It is to be noted that in this case the temperature decreases from the solar interior to upward in the photosphere. The physical mechanism which is causing this variation is largely unknown. However, it appears that the rotation of the solar interior and its atmosphere are linked to show similar variation with temperature.

## ACKNOWLEDGEMENTS

The authors wish to acknowledge the use of data (AIA-94, AIA-131, AIA-171, AIA-193, AIA-211 and AIA-335) from AIA onboard SDO for the period 2012-2018. These are acquired from the webpage of Solar Dynamics Observatory (SDO), a mission of National Aeronautics and Space Administration (NASA). The research at Udaipur Solar Observatory (USO), Physical Research laboratory, Udaipur is supported by Department of Space, Government of India. We also acknowledge the various supports for this research work provided by Department of Physics, Chaudhary Charan Singh University, Meerut, India. We thank the referee for useful comments and suggestions to improve our manuscript.